\documentclass[aps,prd,twocolumn,groupedaddress]{revtex4}

\usepackage{graphicx}
\usepackage{dcolumn}
\usepackage{bm}
\usepackage{amsmath}
\usepackage{amssymb}

\begin{document}


\title{Chaotic orbits for spinning particles in Schwarzschild spacetime}


\author{Chris Verhaaren}
\email[]{chrisverhaaren@gmail.com}
\author{Eric W. Hirschmann}%
\email[]{ehirsch@kepler.byu.edu}
\affiliation{Physics Department, Brigham Young University, Provo, UT 84602}


\date{\today}

\begin{abstract}
We consider the orbits of particles with spin in the Schwarzschild spacetime. 
Using the Papapetrou-Dixon equations of motion for spinning particles, we 
solve for the orbits and focus on those that exhibit chaos using both 
Poincar\'e maps and Lyapunov exponents.  In particular, we develop a 
method for comparing the Lyapunov exponents of chaotic orbits.  We find  
chaotic orbits for smaller spin values than previously thought and find chaotic orbits with astrophysically relevant spin values.
\end{abstract}

\pacs{04.25.D,05.45.Pq}
\keywords{chaos, relativistic orbits, spinning test particles}

\maketitle

\section{\label{sec:intro}Introduction}
A significant amount of effort has gone into gravitational wave detection 
over the last few decades.  Many facilities, including LIGO, VIRGO, and GEO 
are dedicated to their detection and analysis.  In addition, considerable
effort has been made in modeling possible sources for gravitational 
waves.  These include compact binaries with masses on the order of a few solar 
masses.  These detectors are particulaly well tuned to such sources.  However, 
the proposed space based gravitational wave detector, LISA, is better tuned 
to detecting signals such as those coming from compact objects orbiting 
supermassive black holes.  

Such extreme mass ratio inspirals 
(EMRI)~\cite{Tanaka2008} have been studied extensively in recent years.  
In particular, in the test particle limit, the dynamics of these systems, together with 
the inclusion of spin, have been studied in a number of 
works~\cite{Hojman1977,Semerak1999,Suzuki1997,Hartl2002,Hartl2003}.  A subset  
of these studies has considered the question of chaotic orbits in both the 
Schwarzschild~\cite{Suzuki1997} and Kerr~\cite{Hartl2002,Hartl2003} spacetimes.
This work has shown that chaotic orbits are possible in these systems
and is a consequence of the spin orbit coupling.  However, these same studies 
suggest that chaotic orbits exist only when the orbiting 
particle has an unphysically large amount of spin.  
Nonetheless, the parameter space in which to search for chaos in these systems
is large and has not been fully explored. It has been shown for both the Schwarzschild~\cite{Suzuki1999} and Kerr~\cite{Kiuchi2004} spacetimes that chaotic orbits change the character of the energy spectrum of their gravitational waveform. Because chaotic orbits might also lead to chaotic gravitational wave signals, and such signals may be more difficult to detect~\cite{Cornish2001}, the more we can say about these systems the better.

We consider the orbits of spinning test particles in Schwarzschild.  Our 
approach to studying the possibility of chaos in this system is to use 
Poincar\'{e} sections as an indicator of chaos similar to Suzuki and Maeda~\cite{Suzuki1997}. Further, we use and extend a somewhat more sophisticated method 
of calculating the Lyapunov exponent of these orbits as employed by 
Hartl~\cite{Hartl2002,Hartl2003}. As part of this, we present a method for 
extracting an improved prediction of the Lyapunov exponent which allows us to 
distinguish more carefully small Lyapunov exponents from zero. This analysis 
reveals some new classes of chaotic orbits. These chaotic orbits include a previously unsuspected class of orbits that have spins for the inspiraling member that may be obtainable by astrophysical systems.

The remainder of the paper is constructed as follows. We describe the 
equations of motion next together with our choice of supplementary condition. 
In Sections~\ref{sec:chaos} and~\ref{sec:Lyapcomp}, we describe our methods 
for determining the chaos of various orbits, in Section~\ref{sec:results} we present the results of our integrations and conclude in Section~\ref{sec:conclude}.

\section{\label{sec:papa} The Papapetrou-Dixon Equations}
To model a spinning test particle we use the equations of motion of 
Papapetrou~\cite{Papa1951} and Dixon~\cite{Dixon1964}.  These equations 
describe a spinning particle in the pole-dipole approximation.  That is to 
say, they describe the particle as a mass monopole and spin dipole. 
These equations generalize geodesic motion to a spinning particle and 
are written in terms of the momentum, $P^{a}$, and antisymmetric spin tensor,
$S^{ab}$, of the particle.  They can be written as 
\begin{align}
V^{c}\nabla_{c}P^{a}&=-\frac{1}{2}R^{a}_{\phantom{a}bcd}V^{b}S^{cd} \label{eqn:SpinPT}\\
V^{c}\nabla_{c}S^{ab}&=P^{a}V^{b}-P^{b}V^{a}\label{eqn:SpinST}
\end{align}
where $V^{a}$ is the particle's velocity and, again,  
$S^{ab}$ defines the spin of the particle. 
As can be seen from Eq.~(\ref{eqn:SpinST}) the momentum, $P^{a}$, is not 
simply a rescaling of the velocity vector.  Indeed, the momentum is defined as
\begin{equation}
 P^{a}=\mu V^{a}-V_{b}V^{c}\nabla_{c}S^{ab} \label{eqn:Pdef}
\end{equation}
where we will take $\mu$ as the mass of our spinning test particle. 

As shown by Semer\'{a}k~\cite{Semerak1999} this definition of the momentum 
can be combined with Eq.~(\ref{eqn:SpinST}) to solve for $V^{a}$ is terms 
of the momentum and spin of the particle
\begin{equation}
 V^{a}=\frac{\mu}{-P^{b}P_{b}}\left(P^{a}+\frac{2S^{ab}R_{bcde}P^{c}S^{de}}{R_{bcde}S^{bc}S^{de}-4P^{b}P_{b}}\right).\label{eqn:Vdef}
\end{equation}

As they stand, the equations of motions are underdetermined. This is a well 
known issue with these equations and several supplementary conditions have 
been suggested in order to address this 
problem~\cite{Papa1951,Dixon1964,Dixon1970}.  
In this work, we choose to use the supplementary condition $S^{ab}P_{a}=0$. 
This condition, in effect, picks out a center of mass frame for the particle. 
(For more discussion of this and other possible supplementary conditions 
see~\cite{Dixon1970,Ehlers1977}.)  Additionally, this supplementary condition 
implies that the spin tensor, $S^{ab}$, has at most three independent 
components.  As a result, it is possible to reformulate the equations of 
motion in terms of a spin vector, $S^a$, which we will do below.   

As our background spacetime, we will choose spherically symmetric Schwarzschild
spacetime.  This provides a number of conserved quantities with which to 
calculate test particle orbits, even with the assumed spin.   
In particular, it can be shown that for a Killing vector $X^{a}$ the quantity
\begin{equation}
 C=X^{a}P_{a}+\frac{1}{2}S^{bc}\nabla_{b}X_{c}
\end{equation}
is a constant of the motion. As a result, we can define the following  
two constants of the motion 
\begin{align}
 E&=P_{t}+\frac{m}{r^{2}}S^{tr}\\
L&=P_{\phi} - r\sin \theta \left( \sin \theta S^{\phi r}+r\cos \theta S^{\phi \theta}\; \right).  \label{eqn:spinconvL}
\end{align}

In addition to these constants of the motion, the total spin of the particle is 
also conserved.  This quantity, hereafter $S$, is defined as the positive root 
of 
\begin{equation}
 S^{2}=\frac{1}{2}S_{ab}S^{ab}\; .
\end{equation}

In order to understand the physical constraints on $S$, recall that 
lengths are measured in terms of the mass, $m$, of the central object of the 
spacetime and the momentum of the orbiting particle is measured in terms 
of its mass, $\mu$.  We might then think of $S$ as being a unitless 
number multiplying $m\mu$.  Said another way, $S m\mu=l$ where $l$ is the 
spin angular momentum of the particle.  Note that the Papapetrou equations are 
valid in the test particle approximation and therefore only hold for 
$\mu \ll m$.  Because of this, the physical spin of the test particle 
must be much smaller than one in these units.  This can be seen perhaps 
most clearly by 
considering the spinning test particle to be an
extremal Kerr black hole orbiting around a supermassive black hole.
For such an extremal black hole, its angular momentum is $l=\mu^{2}$.   
This leads to a total spin of  
\begin{equation}
S=\frac{l}{m\mu}=\frac{\mu^2}{m\mu}=\frac{\mu}{m}\ll 1\;.
\end{equation}
Using this argument, Hartl~\cite{Hartl2002} estimates physical spins 
as being between about $10^{-4}$ and $10^{-6}$ in $S$.

For numerical simplicity we follow Suzuki and Maeda~\cite{Suzuki1997} and 
Hartl~\cite{Hartl2002,Hartl2003} and modify the equation of motion by working
with the spin vector. This quantity can be defined by
\begin{equation}
  S_{a}=\frac{1}{2}\varepsilon_{abcd}P^{b}S^{cd}\label{eqn:spinvecdef}
\end{equation}
where $\varepsilon_{abcd}$ is the totally antisymmetric tensor density. 
On making the following convenient definitions
\begin{align}
R^{\ast\phantom{ab}cd}_{\phantom{\ast}ab}&= \frac{1}{2}R_{abef}\varepsilon^{efcd}\label{eqn:Rstar}\\
^{\ast}\! R^{\ast abcd}&=\frac{1}{2}\varepsilon^{efab}R^{\ast\phantom{ef} cd}_{\phantom{\ast} ef}\label{eqn:Rstarstar}
\end{align}
the equations of motion\footnote{There are differences in sign with Hartl's 
definitions. The reason is he makes the change of variables $\mu=\cos\theta$ 
which changes the handedness of his coordinate system. This in turn affects 
the overall sign of $\varepsilon_{abcd}$} 
become
\begin{align}
 V^{c}\nabla_{c}P_{a}&=R^{\ast\phantom{ab} cd}_{\phantom{\ast}ab}V^{b}P_{c}S_{d}\\
V^{c}\nabla_{c}S_{a}&=P_{a}\left(R^{\ast b\phantom{c} de}_{\phantom{\ast b} c}S_{b}V^{c}P_{d}S_{e}\right)
\end{align}
With this substitution, the velocity and constants of the motion are now 
defined by
\begin{align}
 V^{a}&=\frac{\mu \left(P^{a}-^{\ast}\!\!\!R^{\ast abcd}S_{b}P_{c}S_{d}\right)}{^{\ast}\! R^{\ast bcde}S_{b}P_{c}S_{d}P_{e}-P^{b}P_{b}}\\
E&=P_{t}+\frac{m}{r^{4}\sin \theta}\left(P_{\theta}S_{\phi}-P_{\phi}S_{\theta}\right)\\
L&=P_{\phi}+\frac{1}{r}\left[ P_{t}\left( r\cos \theta S_{r}-\sin \theta S_{\theta}\right)\right]\nonumber\\
&\phantom{=P_{\phi}}\;+\frac{1}{r}\left[ S_{t}\left(\sin \theta P_{\theta}-r\cos \theta P_{r}\right)\right]\\
S^{2}&=S^{a}S_{a}.
\end{align}

\subsection{\label{Initial Conditions}Initial Conditions}
In order to characterize each orbit in as simple a way as possible, we make 
convenient choices in our initial conditions, working at all times with 
Schwarzschild coordinates.  As every bound orbit will have turning points 
at which the radial velocity will be zero, we choose to begin all orbits 
with $P_{r}=0$.  Also, because the spacetime is spherically symmetric we can 
set $P_{\theta}=0$ and begin motion in the equatorial plane. 

Under these conditions, our normalization, $P^aP_a=-1$, and supplementary 
conditions, $S^{a}P_{a}=0$, reduce to 
\begin{equation}
P_{t}^{2}=\frac{r-2m}{r} + \frac{r-2m}{r^{3}}P_{\phi}^{2}
\end{equation} 
and 
\begin{equation}
S_{t}=\sqrt{\frac{r-2m}{r\,(r^2+ P_{\phi}^{2})}} \, \frac{P_{\phi}S_{\phi}}{r}
\end{equation}
respectively.  These relations allow us to express the total spin as
\begin{equation}
S^{2}=\frac{r-2m}{r}S_{r}^{2}+\frac{1}{r^2}S_{\theta}^{2}+\frac{1}{P_{\phi}^{2}+r^2}S_{\phi}^{2}\;.
\end{equation}

Making the following definitions 
\begin{eqnarray}
\tan\alpha & = \displaystyle{ \frac{S_{\phi}}{S_{r}} } \\
\tan\beta & = - \displaystyle{1 \over S_\theta} \sqrt{S_{r}^{2}+S_{\phi}^{2}} 
\end{eqnarray}
allows us to parametrize the spin vector components in terms of the 
angles $\alpha$ and $\beta$.  These angles are analogous to the $\phi$ and 
$\theta$ of spherical polar coordinates respectively and have their origin at 
the test particle's center of mass.

With these definitions we can specify an orbit by five initial conditions. 
The determining quantities are: $r$, $P_{\phi}$, $S$, $\alpha$, and $\beta$. 
These correspond to the coordinate distance of the test particle from the 
central mass, its momentum in the $\phi$ direction, and the magnitude and 
orientation of its spin. 

\section{\label{sec:chaos}Measuring Chaos}
\begin{figure}[tb]
    \centerline{\includegraphics[height=2.7in, width=3.7in]{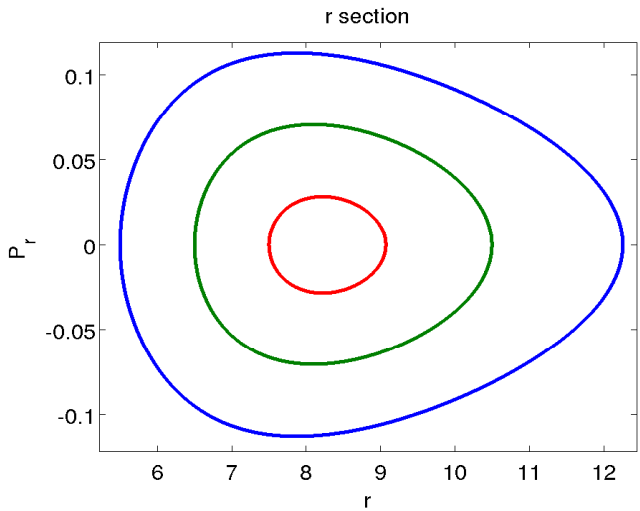}}
    \caption[Three Poincar\'{e} Sections with no chaos]{\label{fig:nesttori}  Three 
Poincar\'{e} sections of the phase space trajectories of nonchaotic particle 
orbits. Each has $L=3.6$ and from the innermost orbit out $E_{1}=0.9499$, $E_{2}=0.9511$, and $E_{3}=0.9534$. Notice that the trajectories are confined to the surface of the tori 
which intersect the section.}
\end{figure}

\begin{figure}[tb]
    \centerline{\includegraphics[height=2.7in, width=3.7in]{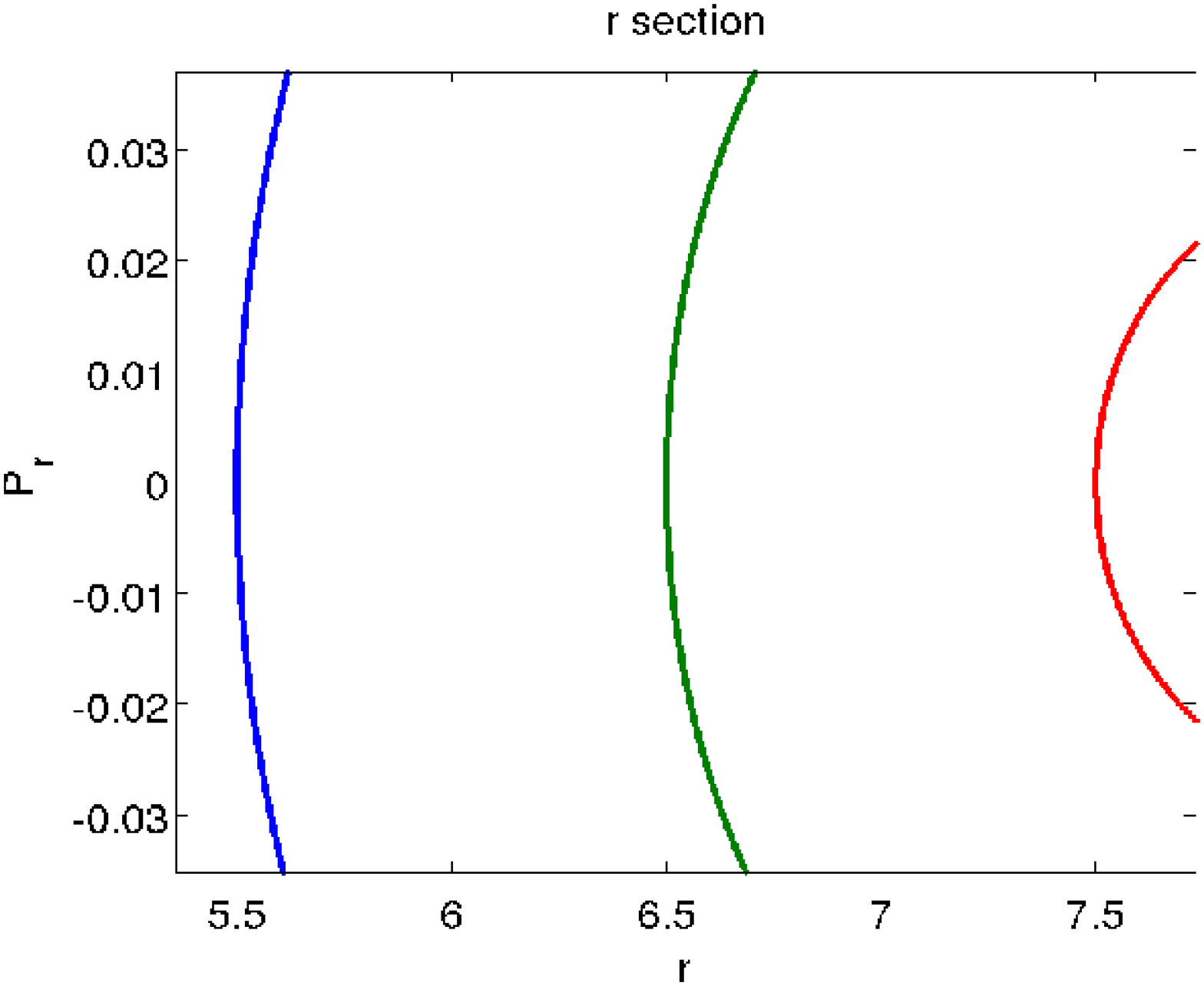}}
    \caption[Three Poincar\'{e} Sections with no chaos]{\label{fig:nesttoriclose}  A close up view of the orbits from Fig. \ref{fig:nesttori}. Notice that this zoomed in view continues to show the trajectories are constrained to the tori. }
\end{figure}

In order to gain confidence in deciding whether a particular orbit is chaotic 
or not we use two tests for chaos.  The first is to check for the 
breaking up of the KAM tori in a Poincar\'{e} section of the phase 
space~\cite{Hilborn,Ott}.  Following Suzuki we choose the section defined 
by the $r-P_{r}$ plane in the phase space, where $r$ is the coordinate 
distance the test particle is from the center of the central mass and $P_{r}$ 
is the conjugate momentum.  In the case where the particle has no spin,  
typical sections look like ovals as in Figure \ref{fig:nesttori}. A closer look at the these plots (Figure \ref{fig:nesttoriclose} ) shows that the phase space trajectories are confined to the surface of a torus.

\begin{figure}[tb]
    \centerline{\includegraphics[height=2.7in, width=3.7in]{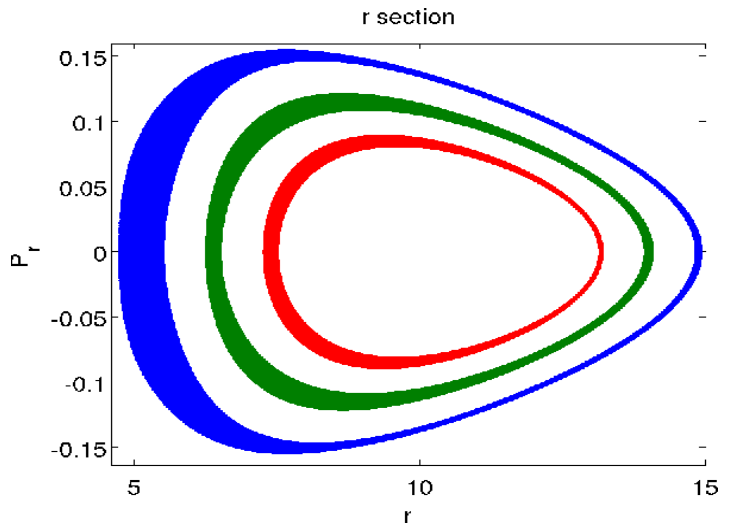}}
    \caption[Poincar\'{e} Section with chaos, example of KAM tori.]{\label{fig:nestchaos}  Three Poincar\'{e} sections of the phase space trajectories of 
chaotic particle orbits. From the inside orbit out the orbits have spin values $S_{1}=1.515$ $S_{2}=1.581$ and $S_{3}=1.627$. Notice that the trajectories stay close to the 
surface of the tori similar to Figure \ref{fig:nesttori}, but do not 
have the thinly defined surface.}
\end{figure}

\begin{figure}[tb]
    \centerline{\includegraphics[height=2.7in, width=3.7in]{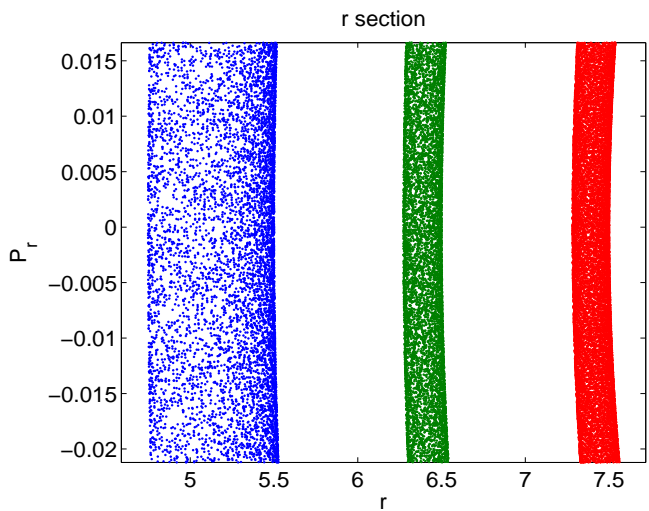}}
    \caption[Poincar\'{e} Section with chaos, example of KAM tori.]{\label{fig:nestchaosclose}  A close up view of the orbits from Fig. \ref{fig:nestchaos}. Notice that this zoomed in view continues to show the trajectories are not constrained to the tori. Compare to Fig. \ref{fig:nesttoriclose}}
\end{figure}

When considering sections for spinning test particle orbits we look for 
this clean oval to break up as in Figure \ref{fig:nestchaos}. A closer look at the bands (Figure \ref{fig:nestchaosclose} ) clearly shows phase space trajectories have left the torus surface. The orbits that produce sections like this are close in to the black hole and are in agreement with the sections produced by~\cite{Suzuki1997}. 

The second method we use to look for chaos in this system is to calculate 
Lyapunov exponents.  If one defines the distance $D$ between two phase 
space trajectories as
\begin{equation}
D\left(t\right)=d_{0}e^{\lambda t}\label{eqn:lyapdef}
\end{equation}
where $d_{0}$ is the initial separation between the two trajectories, then 
$\lambda$ is defined to be the Lyapunov exponent.  When $\lambda$ is greater 
than zero the system is said to be chaotic.  In Hamiltonian systems, of 
which the Papapetrou equations are an example, the Lyapunov exponent can 
never be less than zero.  A negative exponent would indicate some kind of 
attractor in the phase space, but these do not appear in conservative systems. 

In practice, calculating $\lambda$ is a challenge. The Jacobian method 
used by Hartl has the advantage of following just one phase space trajectory. 
The basic idea is that we can track the growth of a vector in the tangent 
space of the trajectory to compute the Lyapunov exponent (for more detail 
see~\cite{Hartl2002} and~\cite{Hartl2003}).  The equation that dictates the 
evolution of this vector is
\begin{equation}
\frac{d\boldsymbol{\xi}}{d\tau}={\bf  Df\cdot \boldsymbol{\xi}}
\end{equation}
where $\boldsymbol{\xi}$ is a vector in the tangent space, and ${\bf Df}$ 
is the Jacobian matrix of the system.

To understand this method, consider a high dimensional ellipsoid in the 
phase space defined by some set of initial conditions.  As the system evolves
away from these initial conditions, this ellipsoid will become warped.   
However, by virtue of Liouville's theorem, the phase space volume of this 
ellipsoid will be conserved.  As a result, any axis of the ellipsoid that 
corresponds to a chaotic coordinate will stretch.  This will necessitate 
that at least one other axis will contract.  The system can have more 
than one chaotic coordinate, but for each chaotic coordinate there must be 
a nonchaotic coordinate whose value converges at the same rate as the 
chaotic coordinate's value diverges.  In other words, for every Lyapunov 
exponent that corresponds to a chaotic axis there is an exponent with the 
same magnitude but opposite sign.

The Jacobian method finds the largest Lyapunov exponent by evolving a vector, 
which is defined by the initial conditions of the system, in the tangent 
space of the phase space.  As the system evolves, this vector lines up with the 
direction of greatest stretching.  By considering the magnitude of this 
vector as a function of the proper time $\tau$, we can then define the 
largest Lyapunov exponent of the system by
\begin{equation}
\lambda=\lim_{\tau \rightarrow \infty} \, \Bigl[ {1 \over \tau} \, \ln \Bigl(\frac{|\boldsymbol{\xi} |}{|\boldsymbol{\xi _{0}}|} \Bigr) \, \Bigr] 
\end{equation}
where $\boldsymbol{\xi _{0}}$ denotes the initial tangent vector.  As
our numerical integrations cannot continue indefinitely, we will, in our 
calculations, refer to the Lyapunov exponent as a function of $\tau$,
\begin{equation}
\lambda(\tau)=\frac{\ln (|\boldsymbol{\xi} |)}{\tau}\label{eqn:lyapvec}
\end{equation}
where we have taken $|\boldsymbol{\xi _{0}}|=1$.

For this last analysis we have denoted the magnitude of a vector 
$\boldsymbol{\xi}$ by $|\boldsymbol{\xi}|$.  Recall that this vector 
lives in the tangent space to the phase space of our physical system.  This 
leads to uncertainty about what norm to use when calculating a vector 
magnitude.  Eckmann~\cite{Eckmann1985} shows that when calculating Lyapunov 
exponents, different norms may lead to different values, but the sign  
of the exponent will not be affected. 
With this in mind, we use the Euclidean norm for simplicity 
when calculating vector magnitudes.

This method can be extended to find the Lyapunov exponent corresponding to 
each axis of the stretching ellipsoid.  Hartl implements this extended 
method~\cite{Hartl2002} in some cases and shows that the Lyapunov 
exponents do come in opposite sign pairs for the spinning particle system. 
His results also indicate that the direction of greatest stretching is not 
in the direction of any one coordinate or conjugate momentum.  Thus, when we 
find the largest exponent we do not expect it to correspond to a particular 
coordinate or momentum.

\section{\label{sec:Lyapcomp}Method for Comparing Lyapunov Exponents}
Because the Jacobian method requires the Lyapunov exponent to be defined in 
terms of a limit we can in practice only approximate its value.  
In previous work, 
a particular orbit was allowed to evolve for some set amount of time and 
with the corresponding value of $\lambda(\tau)$ taken as the approximate 
exponent. 

A problem that arises with this method is that different orbits converge 
to their Lyapunov exponents at different rates.  In particular, the 
Lyapunov exponent for the case $S=0$ approaches zero much more slowly 
than for any other similar orbit with small, nonzero spin.  To address this 
issue, we have developed a 
different method which both reduces computation time and predicts the Lyapunov 
exponent.

\begin{figure}[tb]
    \centerline{\includegraphics[height=2.7in, width=3.7in]{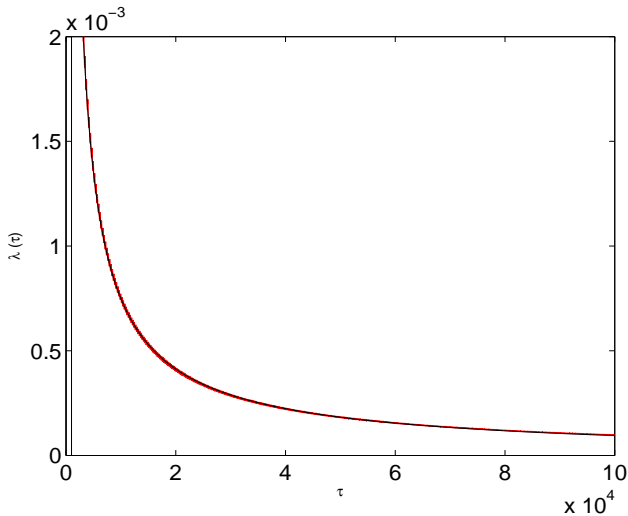}}
    \caption[Curve fitting of Lyapunov exponent in $S=0$ orbit]{\label{fig:lyapfitS0}  We see the jagged Lyapunov function converging to zero for an orbit 
with $S=0$.  We also plot the curve fit to the data.  This orbit begins 
at $r=6m$ with $P_{\phi}=3.6$.  The predicted Lyapunov exponent for this orbit 
is $-1.68\times 10^{-5}$ with an RMS error of is $1.23\times 10^{-5}$.}
\end{figure}

\begin{figure}[tb]
    \centerline{\includegraphics[height=2.7in, width=3.7in]{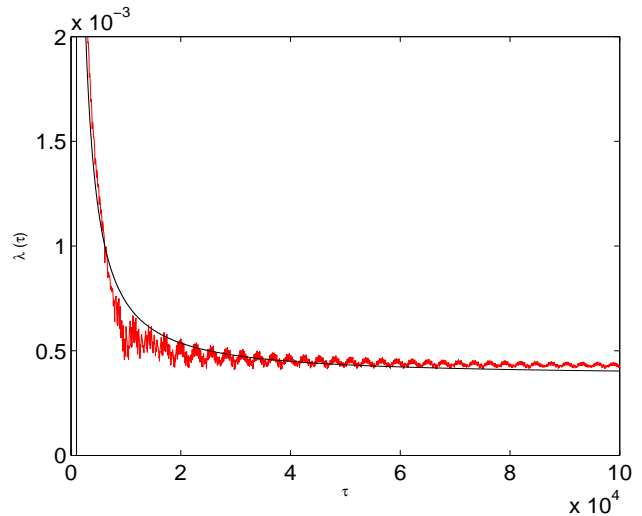}}
    \caption[Curve fitting of Lyapunov exponent in $S=0.5$ orbit]{\label{fig:lyapfitS5} We see the jagged Lyapunov function converging to a nonzero value for an orbit with $S=0.5$. The initial conditions are $r=6m$, $P_{\phi}=3.6$, $\alpha=\frac{\pi}{4}$, and $\beta=\frac{\pi}{4}$. We also plot the curve fit to the data. The predicted Lyapunov exponent is $3.787\times 10^{-4}$ with 0 well outside the RMS error, $6.1\times 10^{-5}$, of the fit.}
\end{figure}

Consider the plots of $\lambda(\tau)$ as shown in Figure \ref{fig:lyapfitS0} 
and Figure \ref{fig:lyapfitS5}.  These are typical examples of how 
$\lambda(\tau)$ converges for a zero spin orbit and a nonzero spin orbit 
respectively.  Notice that the functions are modeled well by the fit
\begin{equation}
f(\tau)=a_{1}+\frac{a_{2}}{\tau ^{a_{3}}}\label{eqn:fit}
\end{equation}
where $a_{1}$, $a_{2}$, and $a_{3}$ are constants that are varied until the 
root mean square error between the fit and $\lambda(\tau)$ is minimized.  More 
explicitly, the constants are varied to minimize 
\begin{equation}
err(\tau)=\sum_{i}^{N}\left(a_{1}+\frac{a_{2}}{\tau_{i} ^{a_{3}}}-\lambda(\tau_{i})\right)^{2}.
\end{equation}
It is then easy to define the RMS error for the fit as
\begin{equation}
RMS=\sqrt{\frac{err}{N}}\;.
\end{equation}

The fitting function sets the Lyapunov exponent of the system to be $a_{1}$. 
Because 
our system is conservative we cannot have negative Lyapunov exponents.  
However, in the orbit corresponding to Figure \ref{fig:lyapfitS0} the root 
mean square error of the fit is $1.23\times 10^{-5}$ while the calculated 
exponent is nonzero and just a bit outside this error range.  Based on other
results, 
such as Figure \ref{fig:lyapfitS5}, as well as  
our method's consistency with the results 
of~\cite{Suzuki1997} and~\cite{Hartl2002}, we are led to belive that 
the fit slightly underestimates the Lyapunov exponent.  
Another example is the orbit corresponding to Figure \ref{fig:lyapfitS5} 
in which the RMS error is $6.1\times 10^{-5}$ keeping 
zero well outside the error bars of the fit.

This model of the Lyapunov function also gives a measure of how quickly the 
exponent converges.  In working with this model we found that orbits with 
large spin had Lyapunov functions that, in general, converged faster than 
those with 
small spin.  It is possible that in previous work similar Lyapunov exponents 
with slower convergence rates might have been discounted as numerical error. 
The 
difference in convergence rates between the zero point and these small spin 
exponents is much smaller than the difference between the zero point and the 
high spin exponents.  If these differing rates of convergence are not 
taken into account, terminating each orbit after some predetermined time 
seems natural.  Unfortunately, our experience shows that this can inflate the 
value of the Lyapunov exponent in the zero spin case.  Because that case 
has the slowest rate of convergence, it can then be 
difficult to distinguish it from cases with nonzero spin and nonzero 
Lyapunov exponents.


We avoid this problem by comparing Lyapunov exponents once a a uniform 
degree of convergence has been reached rather than a particular time. While 
evolving the system we fit the
Lyaponv function to Eq.~(\ref{eqn:fit}) and continue the evolution until 
the derivative of the fitting function reaches a predefined tolerance  
close to zero.  When the derivative of the fit has become sufficiently small, 
specifically when the value is on the order of $10^{-8}$, we say that the 
function has converged.  In this way we compare exponents which have all 
converged the same amount rather than comparing by the overall time of 
evolution.  We 
find that this method reveals more of the chaotic nature of these orbits.

Further, we can use the fit curve to predict the Lyapunov exponent for a 
given orbit without integrating for infinite time.  These predicted values, 
together with an estimate of their error, give us confidence that the true 
exponent is within the corresponding  
range of values and, more importantly, provide a clear differentiation 
between positive and zero values.

\section{\label{sec:results}Results}
A primary motivation for our work is to determine whether or not 
astrophysically relevant spins can give rise to chaotic orbits. We 
first consider orbits 
similar to those considered by Suzuki and Maeda and then look at other 
types of orbits.  Because the orbits considered by Suzuki and Maeda are 
confined to high curvature regions, the additional orbital types we consider 
are those that remain in regions of low curvature and orbits that traverse 
both high and low curvature regions.  For each orbital type we begin with 
a set of initial conditions with $S=0$ and calculate the Lyapunov exponent. 
We then increase the magnitude of the spin and plot the Lyapunov exponents 
for each spin.

\subsection{High Curvature Orbits}
\label{ssec:High Curvature Orbits}
We first consider orbits similar to those used by Suzuki and 
Maeda~\cite{Suzuki1997}.  These orbits remain close to the center of the 
spacetime 
throughout the 
system's evolution. This confines the orbits to regions of high curvature 
which allows the coupling between the curvature and spin to have a large 
effect on the particle's motion.

\begin{figure}[tb]
    \centerline{\includegraphics[height=2.7in, width=3.7in]{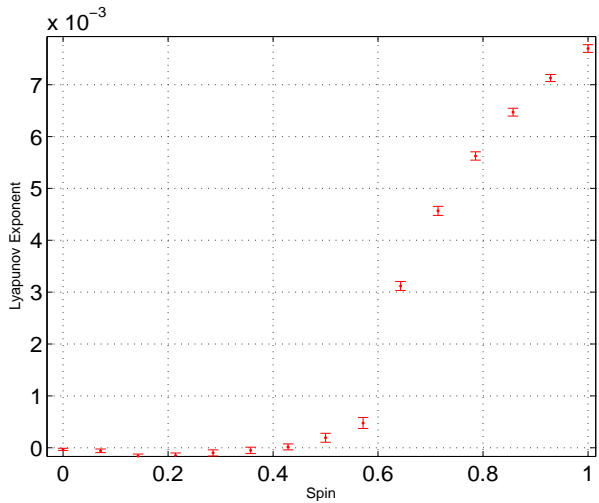}}
    \caption[Lyapunov exponent approximations for a close to center orbit]{\label{fig:fitwitherror} These Lyapunov exponents are given by the constant 
term in the curve fitting model, Eq.~(\ref{eqn:fit}), with root mean 
square error bars.  The initial spin orientation is $\alpha =\frac{\pi}{4}$ 
and $\beta =\frac{\pi}{4}$ and constants of the motion for the spinless case are $L=3.6$ and $E=0.9522$. Compare with Figures \ref{fig:closespindown} 
and \ref{fig:closespinup}. Note that the orbits transition from nonchaotic 
to chaotic around a spin of $S=0.5$.}
\end{figure}

One difference between the orbits we consider and those used by Suzuki and 
Maeda is the orbital angular momentum of the spinning particle.  They consider 
orbits with $L=4.0$ whereas we begin our analysis with orbits having angular 
momentum of $L=3.6$.  Our initial values for energy $E=0.9522$ and an initial 
radius of $r=6m$ are chosen to make the orbit close to circular.  We also keep 
track of the initial conditions in spin that produce these orbits.  In 
particular, we consider what spin magnitudes produce chaotic orbits and how 
the spin vector's initial orientation affects the Lyapunov exponents.  In the 
first case the spin vector's initial orientation is given 
by $\alpha=\frac{\pi}{4}$ and $\beta=\frac{\pi}{4}$. 

\begin{figure}[tb]
    \centerline{\includegraphics[height=2.7in, width=3.7in]{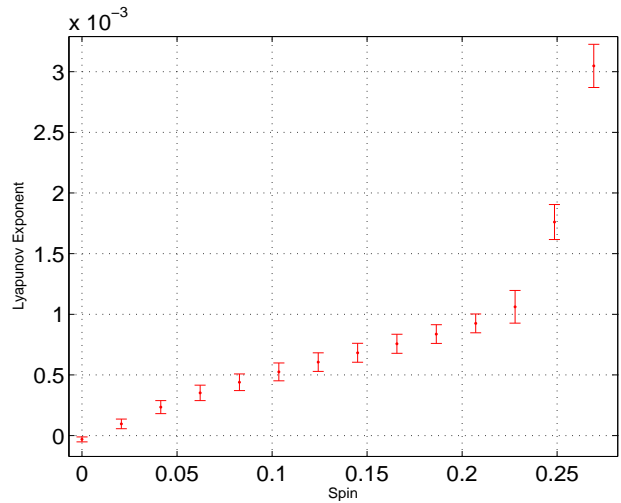}}
    \caption[Lyapunov exponent approximations from curve fitting with $\beta =\pi$.]{\label{fig:closespindown}  These Lyapunov exponents are given 
by the constant value in the curve fitting model with the root mean square 
value providing the error bars.  The initial spin orientation is $\beta =\pi$ and constants of the motion for the spinless case are $L=3.6$ and $E=0.9522$. 
All spin values higher than $S=0.3$ cause the particle to cross the event horizon. Note that 
lower spin values give exponents of the same order of magnitude as 
those in Figure \ref{fig:fitwitherror}.}
\end{figure}

As Figure \ref{fig:fitwitherror} shows, the Lyapunov values for this orbit 
can be put into one of two groups.  From the maximal spin value of one down 
to about $S=0.5$ there seems to be a well defined trend with nonzero Lyapunov 
exponents and corresponding chaotic orbits.  These values are in agreement 
with those reported by Suzuki and Maeda as well as Hartl.  
From $S=0.5$ down to zero spin the orbits are not chaotic.  Again, we find 
excellent agreement with~\cite{Suzuki1997} and~\cite{Hartl2003}.

\begin{figure}[tb]
   \centerline{\includegraphics[height=2.7in, width=3.7in]{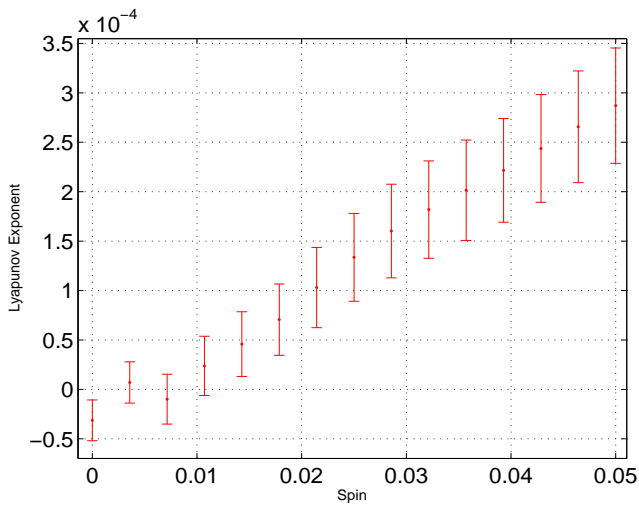}}
    \caption[Lyapunov exponent approximations from curve fitting with $\beta =\pi$ with small spin]{\label{fig:closespindownsml}  These Lyapunov exponents 
correspond to the small spin case of Figure \ref{fig:closespindown} 
with $\beta=\pi$.  Note the monotonically decreasing, yet still positive, 
values of the exponent for decreasing spin values.  This trend continues for 
spins as low as $S=0.015$.  However, this trend  
would appear to be lost for the very smallest values of spin.  Indeed,  
we feel confident to treat these values as zero with correspondingly  
nonchaotic orbits. } 
\end{figure}

The initial spin vector orientation can affect the magnitude of the Lyapunov 
exponent.  As an example, if we change the spin orientation but otherwise 
keep the same initial conditions as before, the particle's behavior changes. 
When $\beta =\pi$, the spin is pointed down perpendicular to the 
equatorial plane, the spin orbit coupling causes the particle to be pulled 
in closer to the center of the spacetime.  We find that when the spin angular 
momentum and orbital angular momentum are parallel there is a repulsive 
spin orbit interaction and when the momenta are antiparallel there is an 
attractive interaction. These results agree with Wald's~\cite{Wald1972} 
analysis of similar systems.  When this interaction becomes strong enough 
the particle will no longer exhibit a bound orbit, and is captured by the 
black hole. In Figure \ref{fig:closespindown} we see the Lyapunov exponent 
values for increasing spin values.  Notice that after $S\approx 0.3$ there 
is no data.  These data points are not included because for $S>0.3$ the 
particle is captured by the black hole.

Notice also that the exponents in Figure \ref{fig:closespindown} 
appear to be nonzero below $S=0.05$.  Since the interaction between the spin 
angular momentum and the orbital angular momentum pulls the particle in 
closer, the particle traverses higher curvature regions of the spacetime 
than the orbits described by Fig~\ref{fig:fitwitherror}.  This spin 
orientation is the only difference between the two cases.  Thus, on comparing 
Figure \ref{fig:fitwitherror} and Figure \ref{fig:closespindown} we can see 
that the orientation of the spin can have a dramatic effect on the dynamics 
of the particle.

\begin{figure}[tb]
    \centerline{\includegraphics[height=2.7in, width=3.7in]{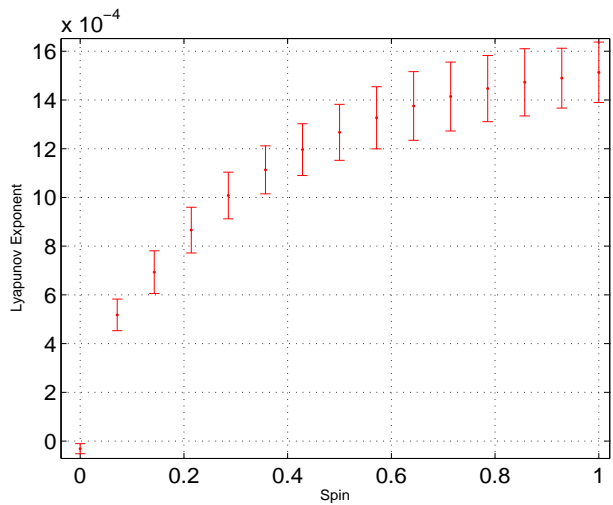}}
    \caption[Lyapunov exponents with $\beta =0$ .]{\label{fig:closespinup}
These Lyapunov exponents are given by the constant value in the curve 
fitting model with the root mean square value providing the error bars. 
The initial spin orientation is $\beta =0$ and constants of the motion for the spinless case are $L=3.6$ and $E=0.9522$. No large exponents 
(compare to Figure \ref{fig:fitwitherror}) appear in this configuration.}
\end{figure}

In Figure \ref{fig:closespindownsml} we zoom in on the small spin value 
orbits.  This data clearly shows positive Lyapunov exponents for spins as 
small as $S\approx 0.015$.  This lower bound in spin is considerably less than
the bounds given in~\cite{Suzuki1997} and ~\cite{Hartl2003}.  Notice that 
exponents in 
Figure \ref{fig:closespindownsml} are small in comparison to 
Figure \ref{fig:fitwitherror}.  Using our particular fitting method here 
to predict the Lyapunov exponent was
crucial in being ablt to distinguish these small exponents from zero.

\begin{figure}[tb]
    \centerline{\includegraphics[height=2.7in, width=3.7in]{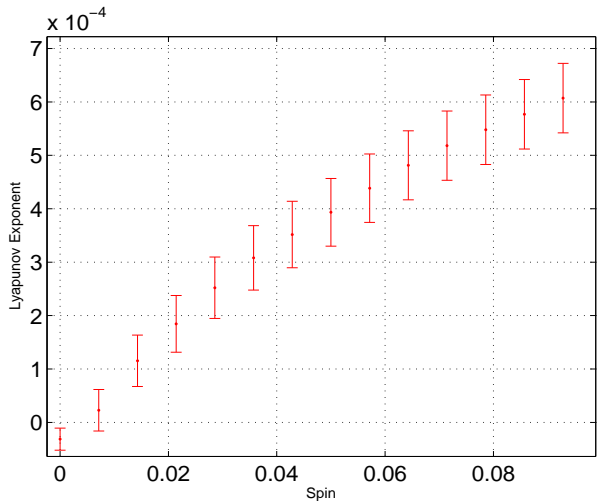}}
    \caption[Lyapunov exponent approximations from curve fitting with $\beta =\pi$]{\label{fig:closespinupsml} These Lyapunov exponents are given by the 
constant value in the curve fitting model with the root mean square providing 
the error bars. The initial spin orientation is $\beta =0$ and constants of the motion for the spinless case are $L=3.6$ and $E=0.9522$. Note that unlike 
Figure \ref{fig:fitwitherror} the exponents all appear to be nonzero.  
No large exponents (compare to Figure \ref{fig:fitwitherror}) appear 
in this configuration. Note the continuous nature to the exponents 
and the apparently positive value for spin as low as $S=0.01$.}
\end{figure}

When we set the initial spin orientation to $\beta =0$, as in 
Figure \ref{fig:closespinup}, the particle is pushed further out from the 
center instead of pulled closer in.  In this case the spin vector and orbital 
angular momentum vector are parallel.  In this case the particle is never 
captured by the black hole. Instead the particle's orbit stays in areas 
of the spacetime with slightly less curvature. Also, the Lyapunov exponents 
stay at the same order of magnitude as the smaller values from 
Figure \ref{fig:fitwitherror} for all spin values, but unlike 
Figure \ref{fig:fitwitherror} no nonzero spins correspond to a zero exponent. 
In Figure \ref{fig:closespinupsml} we look more closely at the small spin 
values for the same orbit. We notice the monotinicity in the predicted exponent 
values, and a nonzero exponent for spin values as low as $S=0.01$.

We also consider orbits which remain in areas of low curvature. 
Choosing nearly circular orbits at radii of $100m$ and calculating the 
Lyapunov exponents as above, we find such orbits to have a uniformly zero 
Lyapunov exponent for every spin orientation which we consider above. 
When the 
radii is sufficiently reduced, $r\approx 19m$, positive exponents are again 
obtained.  These positive exponents first occur for very large spin, but as
the radius continues to decrease, less 
spin is required to achieve a chaotic orbit. 

\subsection{Knife Edge Orbits}
\label{ssec:Knife Edge Orbits}
Knife edge orbits are some of the more interesting test particle 
trajectories allowed in black hole spacetimes.  These orbits have large scale 
precession and execute small tight loops around the center of the system. 
Their dynamics can be understood by considering the effective potential 
in the Schwarzschild spacetime.  For large enough angular momentum this 
potential has a sharp peak close to the center of the spacetime.  These 
``knife edge'' potentials 
are what give these orbits their particular 
dynamics.\footnote{As far as we can tell, this term was first used 
by Ruffini and Wheeler in \emph{The Significance of Space Research for 
Fundamental Physics} in 1970.}  
(For more discussion of these orbits see~\cite{Wheeler}.)  These 
orbits are also known as ``zoom whirl" orbits; the name being used
to describe their behavior, namely  
they zoom in from large radius and whirl about the center. 
(For a comprehensive cataloging of these orbits 
see~\cite{Levin2008}); 

As we consider knife edge orbits for spinning test particles we focus on 
the geometry surrounding their path. Unlike the orbits we have considered 
so far, these move through both high and low curvature regions of the 
spacetime. These orbits start out relatively far from the center of the 
spacetime, 
where the curvature is comparatively low.  But during the course of their 
orbits they execute several small orbits in much higher curvature regions. 
These 
types of orbits help us determine whether chaotic orbits must remain in 
high curvature or just pass through them regularly.

\begin{figure}[tb]
    \centerline{\includegraphics[height=2.7in, width=3.7in]{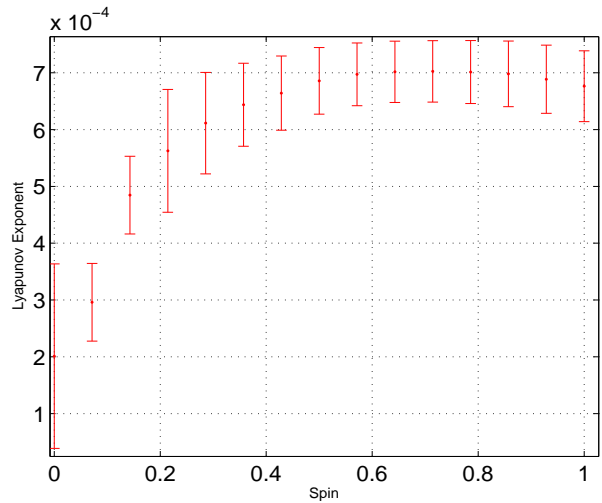}}
   \caption[Lyapunov exponent approximations for a knife edge orbit with $\beta =0$]{\label{fig:knifespin}  These Lyapunov exponents correspond to knife 
edge orbits. Notice that the $S=0$ case is very nonzero.  This is referred 
to as a ``chaos mimic.''  The spinless orbit begins at $r=100m$ and has three 
inner loops at small radius for each large scale orbit.  The initial spin 
orientation of $\beta =0$. In the spinless case $E=0.997$ and $L=3.9246$.}
\end{figure}

In Figure \ref{fig:knifespin} we plot the Lyapunov exponents  
for a knife edge orbit that has an outer radius of $r=100m$ and makes three 
small radius loops for each large radius orbit. The orientation of the 
spin is $\beta =0$.  The angular momentum of the particle is $L=3.9246$ and 
which is nearly the angular momentum of the particles investigated by Suzuki 
and Maeda.  We choose this spin orientation to keep the particle from 
being captured by the black hole.  Recall that this orientation gives an 
effective centrifugal force which pushes the particle away from the black 
hole.  This keeps the particle from being captured, but also reduces the 
number of inner loops traversed in each orbital period.  Thus, as the spin 
increases, the particle can be thought to be retreating from the knife's edge.

\begin{figure}[tb]
   \centerline{\includegraphics[height=2.7in, width=3.7in]{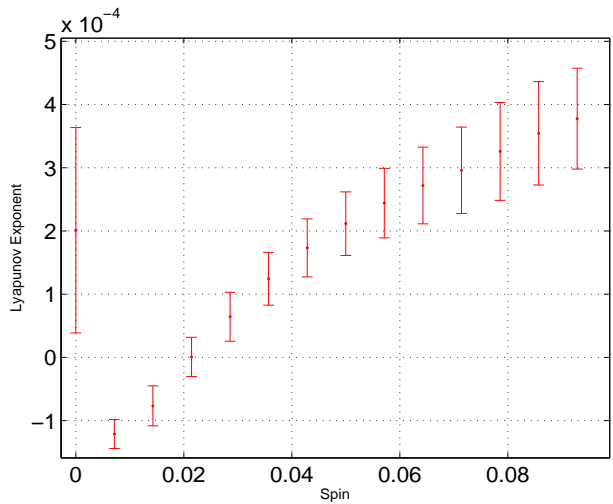}}
   \caption[Small spin value exponent for knife edge orbit with $\beta =0$]{\label{fig:knifespinclose}  These Lyapunov exponents correspond to the same 
knife edge orbits as Figure \ref{fig:knifespin}.  Notice that zero is 
outside the error bars for spins as low as $S=0.03$. }
\end{figure}

One important aspect to notice is that the $S=0$ orbit has a drastically 
nonzero Lyapunov exponent.  This effect is referred to as a 
``chaos mimic''\footnote{These chaos mimics have positive 
Lyapunov exponent due to the dynamics of knife edge orbits. The combination 
of large and small radius orbits causes $\boldsymbol{\xi}$ to grow very 
large.  Interestingly, the effects causing the mimics can be lessened, and 
sometimes removed, by using 
the geodesic equation in contravariant form as the particle's equations of 
motion.  Unfortunately, the chaos mimic that appears in the physical 
chaotic orbit cannot be removed by making this change.}  
and Hartl finds the same effect for knife edge orbits in the Kerr spacetime~\cite{Hartl2003}. This effect casts some doubt on the Lyapunov exponents calculated for the nonzero case.  Some confidence is restored by the trend in the exponents as the spin decreases, specifically that the values approach zero as spin 
goes to zero.  In Figure \ref{fig:knifespinclose} we see a close up look at 
small spin values for the knife edge orbit.  We still have the chaos mimic 
when $S=0$, but we also see a very clear trend in the exponents as the 
spin decreases.

\subsection{\label{ssec:phychaos}Chaotic Orbits With Physical Spin}
\begin{figure}[tb]
    \centerline{\includegraphics[height=2.7in, width=3.7in]{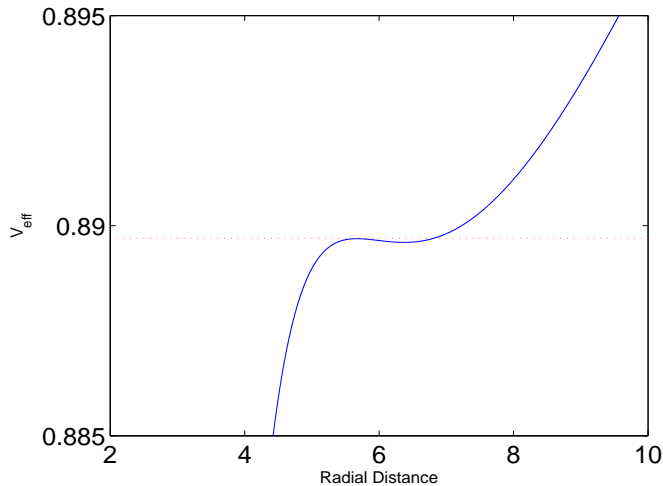}}
    \caption[Effective potential for physical chaotic orbit when $S=0$.]{\label{fig:physpot} This is a plot of the effective potential corresponding to a 
knife edge orbit constrained to high curvature regions. This orbit  
has $L=3.47$ and $E=0.9432$. The line $E^{2}=0.8897$ is shown by the 
dotted line and shows that the particle is constrained to remain near $r=6m$. }
\end{figure}

We now present a particular chaotic orbit at spin values that may be relevant
astrophysically.  In this case, the constants of the motion are $E=0.9432$ and $L=3.47$ and the spin vector orientation is $\beta=\pi$.  These initial conditions define a 
knife edge orbit which is constrained to regions of high curvature.  The 
effective potential for this orbit is shown in Figure \ref{fig:physpot}. 
Notice that the potential well confines the particle with $E^{2}=0.8897$, 
denoted by the dotted line, to remain close to $r=6m$.

\begin{figure}[tb]
    \centerline{\includegraphics[height=2.7in, width=3.7in]{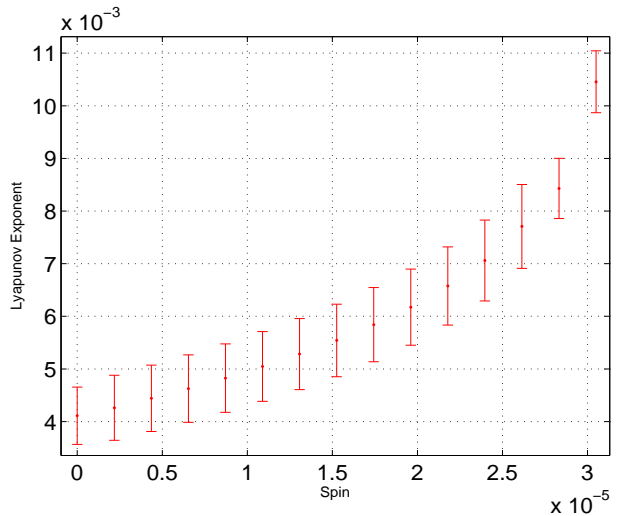}}
    \caption[Lyapunov exponents for physical spin values.]{\label{fig:phychaos} Lyapunov exponents for spin values in the physical range. These orbits have orientation $\beta=\pi$ and constants of the motion for the spinless case are $L=3.47$ and $E=0.9432$.
Note the chaos mimic when $S=0$ and the magnitude of the exponent is 
comparable to the larger values of Figure \ref{fig:fitwitherror}. }
\end{figure}

In Figure \ref{fig:phychaos} we see Lyapunov exponents corresponding to this 
orbit.  Similar to Figure \ref{fig:closespinup}, spin values greater than 
those shown on the graph cause the particle to be captured by the black hole. 
What is more interesting however, is that these nonzero Lyapunov exponents 
correspond to physical spin values and that these values have comparable 
magnitude to the larger exponents of Figure \ref{fig:fitwitherror}.

\begin{figure}[tb]
    \centerline{\includegraphics[height=2.7in, width=3.7in]{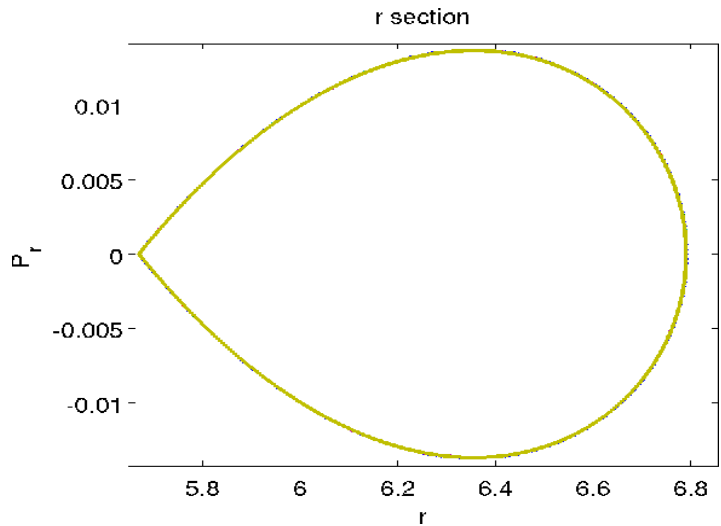}}
    \caption[Poincar\'{e} section in the $r-P_{r}$ plane when $S=0$.]{\label{fig:physsec} These are Poincar\'{e} sections in the $r-P_{r}$ plane for $S=0$ and five orbits with spin magnitudes spaced equally from $S=3.0\times 10^{-5}$ to $S=3.069\times 10^{-5}$. These orbits have the orientation $\beta=\pi$ and constants of the motion for the spinless case are $L=3.47$ and $E=0.9432$. Notice that unlike Figures \ref{fig:nesttori} and \ref{fig:nestchaos} the initial conditions of the orbits make them difficult to distinguish at this level. }
\end{figure}

\begin{figure}[tb]
    \centerline{\includegraphics[height=2.7in, width=3.7in]{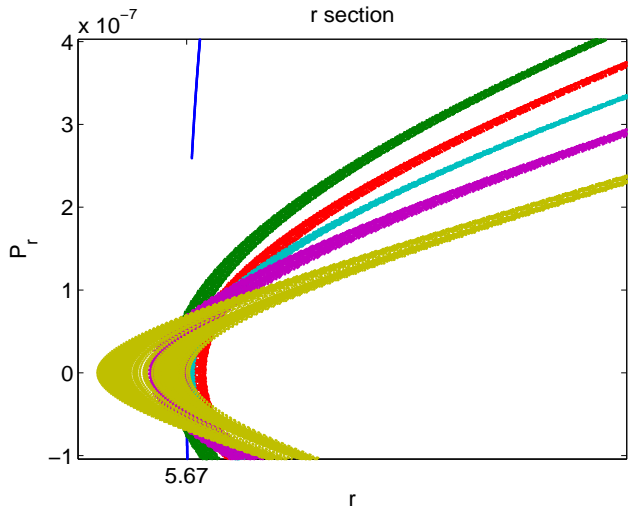}}
    \caption[Poincar\'{e} section in the $r-P_{r}$ plane when $S=3.05\times 10^{-5}$.]{\label{fig:phypoinclose} This is a closer look at Fig. \ref{fig:physsec}. The point $r=5.67$ is marked on the axis which spans $r=5.67-1\times 10^{-6}$ to $r=5.67+3.5\times 10^{-6}$. This finer line in the upper left is the part of the section for the spinless case. Notice the distinct breaking of the tori for the other five orbits. These orbits increase in spin from top to bottom of the upper right corner.}
\end{figure}

One concern with this data is that the exponents seem to converge to the value of the chaos mimic of $S=0$. We can use the KAM tori corresponding to this orbit to resolve this concern. 
In Figures \ref{fig:physsec} and \ref{fig:phypoinclose} we compare the Poincar\'{e} sections of this orbit when $S=0$ and equal steps between $S= 3.0\times 10^{-5}$ and $S=3.069\times 10^{-5}$ respectively. 

Notice that when $S=0$ the phase space trajectory is confined to the surface of the 2 torus intersected by the $r-P_{r}$ plane. We can see this detail in the upper right of Figure \ref{fig:phypoinclose}. In the nonzero spin cases the same Figure shows the breaking up of the tori surfaces which is indicative of chaos. Based on the positive indication for chaos given by both the Lyapunov exponent and the Poincar\'{e} sections we conclude that this orbit is indeed chaotic for some initial conditions that correspond to astrophysically realizable amounts of spin.

As the spin decreases from $S=3.069\times 10^{-5}$ the breaking of the 
torus becomes less and disappears by $2.8\times 10^{-5}$. Thus, it would 
appear that the points of Figure \ref{fig:phychaos} with smaller spins than this amount are not truly chaotic. Since these first points seem to converge to the chaos mimic of the $S=0$ 
case it is likely that the positive exponents corresponding to unbroken tori are inflated by the chaos mimic exactly as in the zero spin case.  We notice from the graph that the final point has smaller error bars than the previous points.  Because this orbit is shown to be chaotic by its Poincar\'{e} section, the true chaotic behavior of the particle may push it 
above the exponent magnitude created by the chaos mimic.

\section{\label{sec:conclude}Conclusion}
We find that chaotic orbits with apparently astrophysically relevant spins 
do exist in the Schwarszchild spacetime.  In particular, we have shown that for a spin 
values between $S=3.0\times 10^{-5}$ and $S=3.069\times 10^{-5}$ both the Lyapunov exponent and the Poincar\'{e} section indicate a chaotic orbit. Recall that Hartl's~\cite{Hartl2003} bounds for chaotic spin values were $10^{-4}$ to $10^{-6}$. These orbits combine the dynamics of knife edge orbits with the high curvature close to the center of the spacetime.  While this is a special class of orbit, decaying orbits may exhibit this behavior before they are captured by the black hole. Indeed, Levin~\cite{Levin2000} has shown 
that compact binaries pass through a chaotic region when inspiraling and 
it is an interesting question as to whether decaying orbits could 
pass through these astrophysical, chaotic orbits. 

The results of our analysis also put the cutoff spin value for chaotic 
orbits much lower than previously thought even for more general orbits.  Suzuki 
and Maeda~\cite{Suzuki1997} give a cutoff value of about $S=0.63$ when the 
orbital angular momentum of the particle is $L=4$.  We have provided strong 
numerical evidence that chaotic 
orbits exist for spin values less than $S=0.01$ for orbits with $L=3.6$. 

Both~\cite{Suzuki1997} and~\cite{Hartl2003} claim that the spin-spin 
interaction in the Kerr metric results in an even greater number of 
chaotic orbits with small spin than in the Schwarzschild spacetime considered 
here.  Indeed, Hartl found a lower bound for spins that yield chaotic orbits
of about $S=0.1$.  As the knife edge orbits that we have investigated here 
also exist in modified form in the Kerr spacetime, we might expect 
that there are even more chaotic orbits with physically relevant spins 
in Kerr.   

A final important question, which is the subject of~\cite{Suzuki1999,Kiuchi2004}, is the potential impact of these chaotic orbits on the gravitational wave emission from these systems.  Certainly, the current results are suggestive that there 
is a chaotic regime in spin which might be astrophysically accessible to 
extreme mass-ratio binaries.  If so, then  
gravitational waves emitted by binary systems with large mass ratios might
contain this chaotic imprint.  
Thus, analyzing these systems for gravitational wave emission 
would be a natural next step in understanding chaos in these systems.

\begin{acknowledgments}
This research was supported in part by NSF grants PHY-0326378 and PHY-0803615 
to BYU and by an ORCA undergraduate mentoring grant from BYU.  
\end{acknowledgments}

\bibliography{paper1}

\end{document}